\documentclass[twocolumn,tighten,times]{aastex61}

\usepackage{natbib}
\usepackage{times}
\usepackage{amsmath}
\usepackage{multirow}
\usepackage{MnSymbol}

\newcommand{\surfdens}{$\Sigma_N(\alpha,\delta)$}

\accepted{by The Astrophysical Journal on 26th March, 2018}
\shorttitle{NGFS Dwarf Galaxies inside half of Fornax's virial radius.}
\shortauthors{Y. Ordenes-Brice\~no et al.}
\begin{document}

\title{The Next Generation Fornax Survey (NGFS): III. Revealing the Spatial Substructure of the Dwarf Galaxy Population inside half of Fornax's Virial Radius.}

\correspondingauthor{Yasna Ordenes-Brice\~no}
\email{yordenes@astro.puc.cl}

\author[0000-0001-7966-7606]{Yasna Ordenes-Brice\~no}
\altaffiliation{PUC-HD Graduate Student Exchange Fellow}
\affiliation{Institute of Astrophysics, Pontificia Universidad Cat\'olica de Chile, Av.~Vicu\~na Mackenna 4860, 7820436 Macul, Santiago, Chile}
\affiliation{Astronomisches Rechen-Institut, Zentrum f\"ur Astronomie der Universit\"at Heidelberg, M\"onchhofstra{\ss}e 12-14, D-69120 Heidelberg, Germany}

\author[0000-0001-8654-0101]{Paul~Eigenthaler}
\altaffiliation{CASSACA Postdoctoral Fellow}
\affiliation{Institute of Astrophysics, Pontificia Universidad Cat\'olica de Chile, Av.~Vicu\~na Mackenna 4860, 7820436 Macul, Santiago, Chile}
\affiliation{Chinese Academy of Sciences South America Center for Astronomy and China-Chile Joint Center for Astronomy, Camino El Observatorio 1515, Las Condes, Santiago, Chile}

\author[0000-0003-3009-4928]{Matthew A.~Taylor}
\altaffiliation{Gemini Science Fellow}
\affiliation{Gemini Observatory, Northern Operations Center, 670 North A'ohoku Place, Hilo, HI 96720, USA}

\author[0000-0003-0350-7061]{Thomas H.~Puzia}
\affiliation{Institute of Astrophysics, Pontificia Universidad Cat\'olica de Chile, Av.~Vicu\~na Mackenna 4860, 7820436 Macul, Santiago, Chile}

\author[0000-0002-5897-7813]{Karla~Alamo-Mart\'inez}
\altaffiliation{FONDECYT Postdoctoral Fellow}
\affiliation{Institute of Astrophysics, Pontificia Universidad Cat\'olica de Chile, Av.~Vicu\~na Mackenna 4860, 7820436 Macul, Santiago, Chile}

\author[0000-0002-3004-4317]{Karen X.~Ribbeck}
\affiliation{Institute of Astrophysics, Pontificia Universidad Cat\'olica de Chile, Av.~Vicu\~na Mackenna 4860, 7820436 Macul, Santiago, Chile}

\author[0000-0003-1743-0456]{Roberto~P.~Mu\~noz}
\affiliation{Institute of Astrophysics, Pontificia Universidad Cat\'olica de Chile, Av.~Vicu\~na Mackenna 4860, 7820436 Macul, Santiago, Chile}

\author[0000-0003-1632-2541]{Hongxin Zhang}
\altaffiliation{FONDECYT Postdoctoral Fellow}
\affiliation{Institute of Astrophysics, Pontificia Universidad Cat\'olica de Chile, Av.~Vicu\~na Mackenna 4860, 7820436 Macul, Santiago, Chile}
\affiliation{Chinese Academy of Sciences South America Center for Astronomy and China-Chile Joint Center for Astronomy, Camino El Observatorio 1515, Las Condes, Santiago, Chile}

\author[0000-0002-1891-3794]{Eva K.\ Grebel}
\affiliation{Astronomisches Rechen-Institut, Zentrum f\"ur Astronomie der Universit\"at Heidelberg, M\"onchhofstra{\ss}e 12-14, D-69120 Heidelberg, Germany}

\author[0000-0002-5322-9189]{Sim\'on~\'Angel}
\affiliation{Institute of Astrophysics, Pontificia Universidad Cat\'olica de Chile, Av.~Vicu\~na Mackenna 4860, 7820436 Macul, Santiago, Chile}

\author[0000-0003-1184-8114]{Patrick C{\^o}t{\'e}}
\affiliation{NRC Herzberg Astronomy and Astrophysics, 5071 West Saanich Road, Victoria, BC V9E 2E7, Canada}

\author[0000-0002-8224-1128]{Laura Ferrarese}
\affiliation{NRC Herzberg Astronomy and Astrophysics, 5071 West Saanich Road, Victoria, BC V9E 2E7, Canada}

\author[0000-0002-2363-5522]{Michael Hilker}
\affiliation{European Southern Observatory, Karl-Schwarzchild-Str. 2, D-85748 Garching, Germany}

\author[0000-0002-7214-8296]{Ariane~Lan\c{c}on}
\affiliation{Observatoire astronomique de Strasbourg, Universit\'e de Strasbourg, CNRS, UMR 7550, 11 rue de l'Universite, F-67000 Strasbourg, France}

\author[0000-0003-4197-4621]{Steffen Mieske}
\affiliation{European Southern Observatory, 3107 Alonso de C\'ordova, Vitacura, Santiago}

\author[0000-0002-5665-376X]{Bryan W.~Miller}
\affiliation{Gemini Observatory, South Operations Center, Casilla 603, La Serena, Chile}

\author[0000-0002-2204-6558]{Yu Rong}
\altaffiliation{CASSACA Postdoctoral Fellow}
\affiliation{Institute of Astrophysics, Pontificia Universidad Cat\'olica de Chile, Av.~Vicu\~na Mackenna 4860, 7820436 Macul, Santiago, Chile}
\affiliation{Chinese Academy of Sciences South America Center for Astronomy and China-Chile Joint Center for Astronomy, Camino El Observatorio 1515, Las Condes, Santiago, Chile}

\author[0000-0003-4945-0056]{Ruben S\'anchez-Janssen}
\affiliation{STFC UK Astronomy Technology Centre, Royal Observatory, Blackford Hill, Edinburgh, EH9 3HJ, UK}

%% Note that the \and command from previous versions of AASTeX is now
%% depreciated in this version as it is no longer necessary. AASTeX 
%% automatically takes care of all commas and "and"s between authors names.

%% AASTeX 6.1 has the new \collaboration and \nocollaboration commands to
%% provide the collaboration status of a group of authors. These commands 
%% can be used either before or after the list of corresponding authors. The
%% argument for \collaboration is the collaboration identifier. Authors are
%% encouraged to surround collaboration identifiers with ()s. The 
%% \nocollaboration command takes no argument and exists to indicate that
%% the nearby authors are not part of surrounding collaborations.

%% Mark off the abstract in the ``abstract'' environment. 
\begin{abstract}
We report the discovery of 271 previously undetected dwarf galaxies in the outer Fornax cluster regions at radii $r_{\rm vir}/4<\!r\!<r_{\rm vir}/2$ using data from the {\it Next Generation Fornax Survey}\,(NGFS) with deep coadded $u'$, $g'$ and $i'$ images obtained with Blanco/DECam at Cerro Tololo Interamerican Observatory.~From the 271 dwarf candidates we find 39 to be nucleated.~Together with our previous study of the central Fornax region, the new dwarfs detected with NGFS data are 392, of which 56 are nucleated.~The total Fornax dwarf galaxy population from NGFS and other catalogs rises, therefore, to a total of 643 with 181 being nucleated, yielding an overall nucleation fraction of $28\%$.~The absolute $i'$-band magnitudes for the outer NGFS dwarfs are in the range $-18.80\le\,M_{i'}\le\,-8.78$ with effective radii $r_{{\rm eff},i'}\,=\,0.18-2.22$\,kpc and an average Sersic index $\langle n\rangle_{i'}\,=\,0.81$.~Non-nucleated dwarfs are found to be fainter and smaller by $\Delta\langle M_{i'}\rangle\!=\!2.25$\,mag and $\Delta\langle r_{{\rm eff},i'}\rangle\!=\!0.4$\,kpc than the nucleated dwarfs.~We demonstrate a significant clustering of dwarf galaxies on scales $\lesssim\!100$\,kpc, and projected surface number density profile estimates, $\Sigma_N(r)$, show a concentration of dwarfs in the Fornax core region within $r\!\la\!350$\,kpc.~$\Sigma_N(r)$ has a flat distribution up to $\sim\!350$\,kpc, beyond which it declines for the non-nucleated dwarfs.~The nucleated dwarfs have a steeper $\Sigma_N(r)$ distribution, being more concentrated towards NGC\,1399 and decreasing rapidly outwards.~This is the first time the transition from cluster to field environment has been established for the very faint dwarf galaxy population with robust sample statistics.

\end{abstract}

\keywords{galaxies: clusters: individual (Fornax) --- galaxies: dwarf --- galaxies: elliptical and lenticular, cD}

%%%%%%%%%%%%%%%%%%%%%%%%%%%%%%%%%%%%
%%%%%%%%%%%%%%%%%%%%%%%%%%%%%%%%%%%%
%%%%%%%%%%%%%%%%%%%%%%%%%%%%%%%%%%%%
\section{Introduction} \label{sec:intro}
Large numbers of faint, low surface brightness dwarf galaxies are rapidly being discovered in different environments throughout the local Universe \citep[e.g.,][]{vanDokkum15,mun15,mul17a,Wittmann17}.~The rise of large detector arrays in present-day observatories such as the Dark Energy Camera \citep[DECam;][]{fla15}, enables us to survey large areas of the sky down to ultra low surface brightness levels, providing the exciting opportunity to search for undiscovered faint dwarf galaxies in massive clusters, galaxy groups, and in the field \citep[e.g.,][]{mun15, fer16, mul17a, ven17}.~Dwarf galaxies are found in galaxy group and cluster environments, which are numerically dominated by early-type dwarf galaxies with characteristically smooth morphologies, exponential surface brightness profiles, and stellar populations consistent with red-sequence galaxies \citep{sandage84, bin85, con03, hil08, dbr11, ord16, Roediger17, eigen18}.~Early-type dwarf galaxies have typically been classified as dwarf ellipticals (dE), but are also known as dwarf spheroidals (dSph) at fainter magnitudes \citep{gre03}.~They exhibit absolute $B$-band magnitudes fainter than $M_B \simeq -16$, corresponding to $\log({\cal M}_{\star}/M_\odot)\!\la\!9$, and effective radii smaller than $\sim\!1$\,kpc \citep[cf.~Fig.~8 in][]{eigen18}.~A further morphological distinction among dwarf galaxies is whether or not they host a central nuclear star cluster.~Recent findings show that dwarf nucleation probability is strongly dependent on its spheroid luminosity \citep{mun15,yob18}, with the fraction of nucleated dwarfs systematically increasing toward brighter magnitudes.

Large populations of low-mass dwarf galaxies are ideal for studying the dependence of galaxy formation and evolution processes in the transition zones between field and cluster environments, especially in rich galaxy clusters.~Statistically significant samples allow us to study their clustering properties on large scales \citep{mun15} and potentially probe the dark matter (DM) fine-structure within the cluster halo.~This distribution in return serves as an ideal laboratory for comparisons with predictions from structure formation models \citep[e.g.][]{bov16}.

Recently, observations have revealed that low-mass dwarf galaxies appear to form surprisingly thin planes in the local Universe \citep{Palowski12, Ibata13, Tully15}.~The observations of such planes challenge current $\Lambda$CDM models of hierarchical structure formation \citep{Kroupa05, Pawlowski14}.~The frequency of the occurrence of such anisotropic distributions in dense environments, will inform an updated view of structure formation in modern galaxy formation models.~These findings show that deep homogeneous surveys are necessary to highlight the spatial distribution of the faint dwarf galaxies in nearby groups and clusters of galaxies without completeness.

In the present work we attempt to lay the foundation for addressing these issues by investigating the faint dwarf galaxy population in the Fornax galaxy cluster out to half of its virial radius ($r_{\rm vir}$), using data obtained as part of the {\it Next Generation Fornax Survey} (NGFS).~The most prominent survey covering the galaxy cluster outskirts \citep{Ferg89}, while seminal, is showing signs of age and suffers from shallow detection limits ($m_B\!\lesssim\!20$\,mag for point-sources and $\mu_B\!\lesssim\!24$\,mag arcsec$^{-2}$ in surface-brightness sensitivity) compared to the potential of modern instrumentation.~Given the faint surface brightness of dwarf galaxies recently detected in various galaxy aggregates, the goal of this paper is to update the known population of Fornax dwarf galaxies out to $\la\!r_{\rm vir}/2$, where one might expect to witness the transition from the central galaxy population to those residing in the cluster outskirts.~The Fornax cluster is the nearest high-density region in the Southern hemisphere \citep[$m\!-\!M\!=\!31.51$\,mag or $D_L\!=\!20.0$\,Mpc,][]{bla09}, and given its proximity, twice the central galaxy density, and a larger early-type galaxy (ETG) fraction than its Northern hemisphere counterpart the Virgo Cluster, Fornax is an important nearby laboratory to investigate the dependence of galaxy evolution on the dynamical state of the environment.

\begin{figure*}[!t]
\centering
%trim=left bottom right top
\includegraphics[trim=7.4cm 0.8cm 7.cm 1.cm,clip,width=0.1345\textwidth]{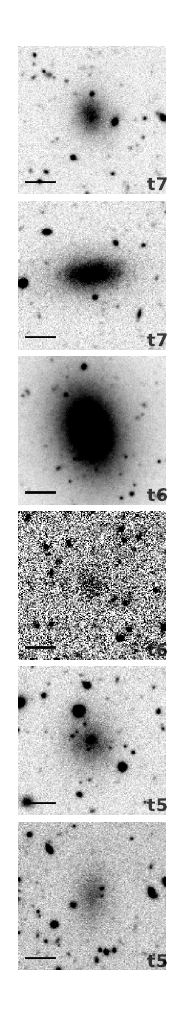}
\includegraphics[trim=0.25cm 0.14cm 0.35cm 0.2cm,clip,width=0.722\textwidth]{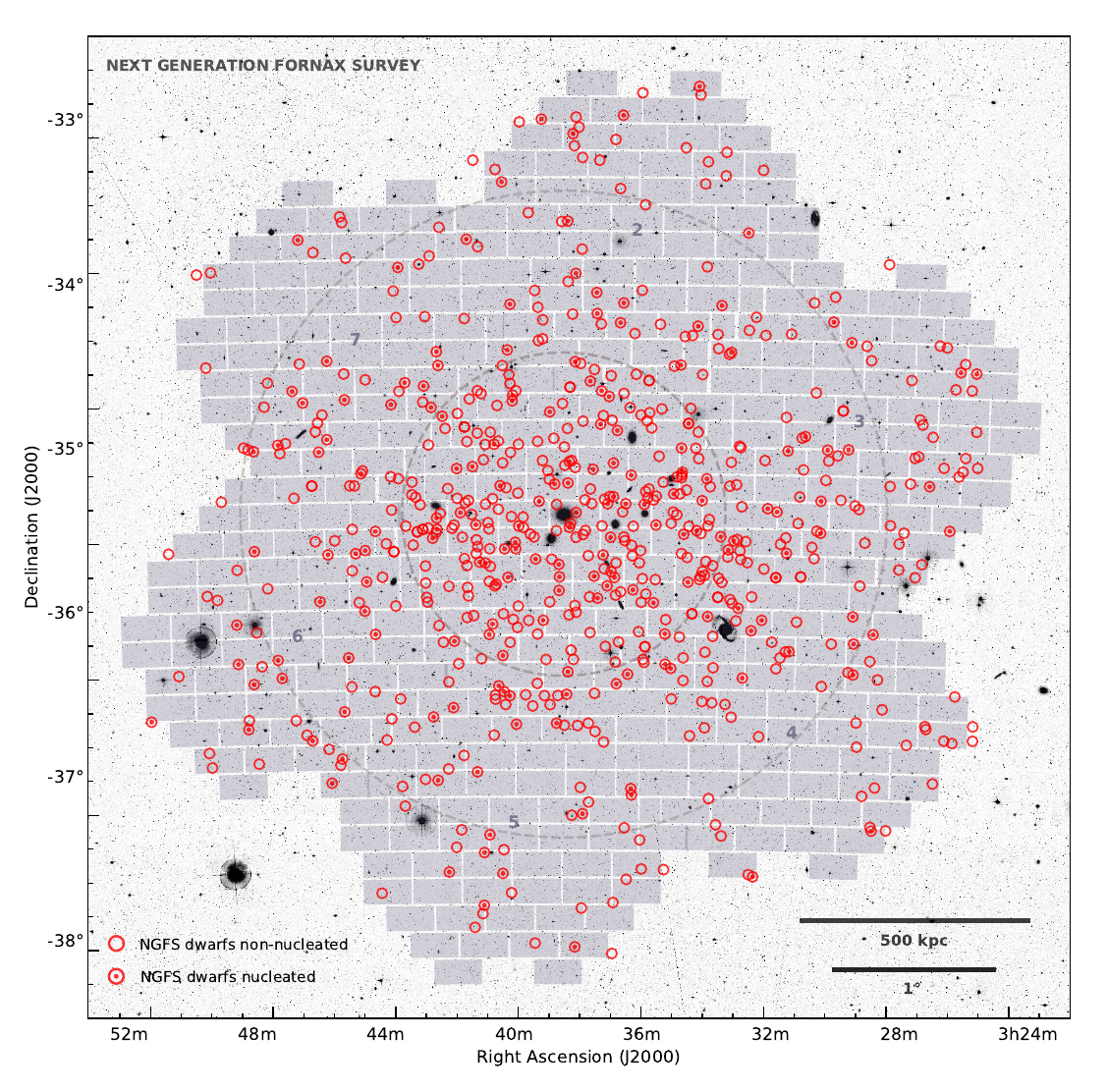}
\includegraphics[trim=7.4cm 0.8cm 7cm 1.cm,clip,width=0.1345\textwidth]{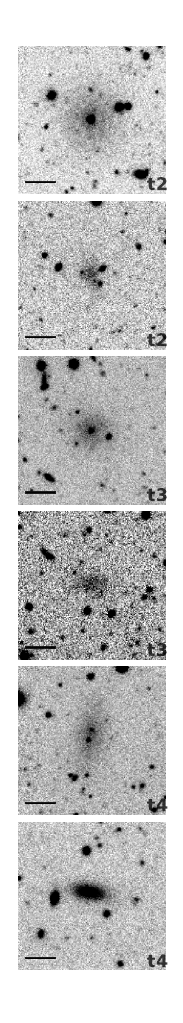}
\caption{Illustration of the spatial distribution of Fornax dwarf galaxy candidates in a greyscale underlying image from DSS.~The footprint of the NGFS survey (inside $\sim\!r_{\rm vir}/2$), which is indicated by the grey shaded DECam tiles with tile\,1 for the central region and tiles\,2-7 for the outer region \citep[see also NGFS footprint in Fig.\,1][]{eigen18}, labeled with their respective numbers (middle panel).~The position of non-nucleated dwarfs are shown by red open circles while the nucleated dwarfs are marked by red circles with a central dot.~Left and right panels show examples of the newly detected dwarfs in the different NGFS footprint tiles (2-7).~Each stamp image contains the NGFS tile number (bottom right) which contains the corresponding dwarfs galaxy as well as  the scale bar (solid line) showing $10.3\arcsec\,\hat{=}\,1$\,kpc at the bottom left of each panel.~Gray-dashed circles show NGC\,1399-centric distances of $r_{\rm vir}/4$ ($\simeq\!350$\,kpc) and $r_{\rm vir}/2$ ($\simeq\!700$\,kpc).}
\label{fig:footprint}
\end{figure*}

%%%%%%%%%%%%%%%%%%%%%%%%%%%%%%%%%%%%
%%%%%%%%%%%%%%%%%%%%%%%%%%%%%%%%%%%%
%%%%%%%%%%%%%%%%%%%%%%%%%%%%%%%%%%%%
\section{Observations and image processing}
\label{sec:observations}
The data presented in this paper is part of the observed {\it Next Generation Fornax Survey} \citep[NGFS;][]{mun15}, an ongoing, panchromatic $\sim\!30\,{\rm deg}^2$ survey of the Fornax galaxy cluster using the {\it Dark Energy Camera} \citep[DECam;][]{fla15} mounted on the 4m Blanco telescope at Cerro Tololo Interamerican Observatory (CTIO).~Figure~\ref{fig:footprint} illustrates the Fornax cluster region that is covered by our inner NGFS footprint, which consists of seven tiles centered on the dominant NGC\,1399 galaxy, and homogeneously mapping the cluster out to $\sim\!50$\% of its virial radius \citep[$r_{\rm vir}\!\simeq\!1.4$\,Mpc;][]{dri01}.~The dwarf galaxy population in the central tile corresponding to $r \leq r_{\rm vir}/4$ was already studied in \cite{mun15} and \cite{eigen18}.~All tiles are observed in three optical bands reaching point-source detections with ${\rm S/N}\ga5$ at 26.5, 26.1, and 25.3\,AB mag in the $u'$-, $g'$-, and $i'$-band, respectively.

Initial image processing is carried out by the CTIO Community Pipeline \citep[CP;][]{val14}, focusing mainly on instrumental signature removal (e.g., bias subtraction, flat-fielding, cross-talk correction).~After the CP-processing, we apply further processing using the {\sc Astromatic}\footnote{www.astromatic.net/software} software suite to astrometrically calibrate and stack the individual frames, and conduct basic photometry \citep[{\sc scamp}, {\sc swarp}, {\sc source extractor}, hereafter SE;][]{ber96,ber02,ber06}.~Astrometric and photometric calibration has been performed using the 2MASS astrometric Point Source Catalogue \citep[][]{skr06} and SDSS $u'g'i'$ stripe 82 standard stars, respectively.~To ensure accurate photometric calibration, we cross-verified it with the globular cluster catalog from \cite{kim13} in the Fornax area, which was compiled using $U,B,V$ and $I$-band photometry taken with MOSAIC-II camera on CTIO/Blanco.~We use the empirical transformation equations from \cite{jordi06} for the comparison, finding good agreement within the uncertainties.~Finally, we proceed to the final image stacking and source detection.

\begin{deluxetable*}{cccccccccc}
\tablecaption{Dwarf galaxies in the Fornax Cluster\label{tab:dwarftable}}
%-------------------------------------------------------------------------------------------------------------------------------------------------------------------------------------------------------------------------------------------------------------------------------------------------------------------------------------------------------------------------------------------------------------------
\tablehead{                                                                                                                                                                                                                                                                                                                               
\colhead{\multirow{2}{*}{ID}}       &   \colhead{\multirow{2}{*}{$\alpha_{2000}$}} &   \colhead{\multirow{2}{*}{$\delta_{2000}$}} &   \colhead{$m_i$}                 &  \colhead{$M_{i}\,$\tablenotemark{a}}     &  \colhead{\multirow{2}{*}{$n$\tablenotemark{b}}}   &  \colhead{${r_{{\rm{eff}}}}$}  &  \colhead{${r_{{\rm{eff}}}}$\tablenotemark{a}}   &   \colhead{\multirow{2}{*}{Type\tablenotemark{c}}}  &   \colhead{\multirow{2}{*}{Reference}} \\      
                                    &                                              &   \                                          &   \colhead{[mag]}               &  \colhead{[mag]}                          &                                   &  \colhead{[arcsec]}            &  \colhead{[kpc]}                &                                       &                                                 }           
%-------------------------------------------------------------------------------------------------------------------------------------------------------------------------------------------------------------------------------------------------------------------------------------------------------------------------------------------------------------------------------------------------------------------
\startdata  
%NGFS033239-332028 &  03:32:39.20 &  $-$33:20:28.4 &  19.36 & $-$12.15 &   0.9968 &  4.06  & 0.394  &  $\medcircle$ & \\
NGFS033260-341909 &  03:32:59.92 &  $-$34:19:08.8 &  20.07 & $-$11.44 &   1.3208 &  5.13  & 0.497  &  $\medcircle$ & \\
NGFS033304-334329 &  03:33:03.61 &  $-$33:43:29.2 &  15.78 & $-$15.73 &   1.2081 &  12.90 &  1.251 &  $\odot$  &  FCC111 \\
\enddata
\tablenotetext{a}{Assuming a distance modulus of $(m\!-\!M)_0\!=\!31.51$ mag or $D_L\!=\!20.0$\,Mpc \citep{bla09}.}
\tablenotetext{b}{S\'{e}rsic index \citep{ser63, cao93}.}
\tablenotetext{c}{Morphological galaxy type classification: $\odot$=nucleated, $\medcircle$=non-nucleated dwarf galaxy.}
\tablecomments{Table~\ref{tab:dwarftable} is published in its entirety in the electronic edition of the {\it Astrophysical Journal}.~A portion is shown here for guidance regarding its form and content.}
\end{deluxetable*}

In previous contributions \citep[][]{mun15,eigen18}, we presented the NGFS results focusing on the inner $\sim\!3\,{\rm deg}^2$ ($\la\!r_{\rm vir}/4$) region of the Fornax cluster.~We presented optical colors and structural parameters for $\ga\!250$ dwarf galaxy candidates, reaching out to a projected distance of $\sim\!350$\,kpc from NGC\,1399, with the spatial distribution of which suggesting a rich and sub-structured dwarf galaxy population extending well beyond these limits.

In this paper we expand upon these initial results by using the surrounding DECam pointings (NGFS tiles 2-7; see Fig.\,\ref{fig:footprint}) to identify 271 new dwarf galaxy candidates, of which 39 are nucleated.~In the following, we refer to the surrounding NGFS tiles (2-7) as the outer footprint.~Together with the 121 previously found NGFS dwarfs from the central tile \citep{mun15, eigen18}, there is a total of 392 new dwarfs, out of which 56 are nucleated, that were discovered in the NGFS data.~Our by-eye dwarf detection strategy used for the central tile is again utilized, where several members of the NGFS team (KAM, KXR, MAT, PE, THP, YOB) independently compiled dwarf candidate lists for tiles 2-7 using RGB full-color image stacks, constructed from the individual $u'g'i'$ frames.~Cross-matching lists and setting a minimum threshold of three independent detections yields a new robust list of dwarf galaxy candidates projected within $r_{\rm vir}/2$ of NGC\,1399.~While all frames are fully reduced, we defer a full color and stellar population analysis to a future paper and limit the scope of the present work to a monochromatic $i'$-band presentation of magnitudes, structural parameters and spatial distribution characteristics.

We complement our final catalog with the known dwarf galaxy population in Fornax, using the likely members in the Fornax Cluster Catalog \citep[FCC,][]{Ferg89}.~In the outer NGFS footprint ($0.25\!<\!R/R_{\rm vir}\!\leq\!0.5)$, we find a total of 114 FCC galaxies ($29.6\%$) and in the so-far searched NGFS survey area a total of 251 literature galaxies \citep[$39\%$;][]{Ferg89, Mieske07}.~Taking into account the dwarfs from existing catalogs and the new NGFS dwarfs, the total Fornax dwarf galaxy population reported in this work consists of 643 dwarfs, of which 462 are non-nucleated and 181 are nucleated.

%%%%%%%%%%%%%%%%%%%%%%%%%%%%%%%%%%%%
%%%%%%%%%%%%%%%%%%%%%%%%%%%%%%%%%%%%
%%%%%%%%%%%%%%%%%%%%%%%%%%%%%%%%%%%%
\section{Structural parameters of the dwarf candidates}
The surface brightness profiles for the dwarfs are studied with GALFIT \citep[v3.0.5][]{peng02} using a S\'ersic profile \citep{ser63,cao93}.~Our procedure has been described in detail in \cite{mun15} and \cite{eigen18}.~It is an iterative process where the light profile is approximated with a one-component fit to the 2D galaxy surface brightness distribution.~We run GALFIT on cutout images of $105''\!\times\!105''$ in size ($\hateq$\,10.2\,kpc\,$\times$\,10.2\,kpc) using object masks created from SE segmentation maps and PSF models created with {\sc PSFex} \citep{ber11}.~For nucleated dwarfs, we iterate the method described above several times and improve the object mask in each iteration step until the nucleus is completely masked \citep[see][for more details]{eigen18, yob18}.~The final fitting profile considers only the spheroid light component of the dwarf galaxy, leaving a residual image (original\,--\,spheroid model) with the nuclear cluster in the galaxy center.~The analysis of the nuclear star clusters will be presented in an upcoming NGFS contribution.~In the following we focus on the structural parameters of the spheroid sample.~The dwarf candidates from the outer footprint have absolute $i'$-band magnitudes in the range $-18.80 \le M_{i'} \le -8.78$ with photometric errors $<\!0.1$\,mag, effective radii between $1.8''$ to $22.8''$ ($r_{{\rm eff},i'}\!=\!0.18\!-\!2.22$\,kpc at the Fornax cluster distance of $D_L\!=\!20.0$\,Mpc), a mean S\'ersic index of $\langle n\rangle_i\!=\!0.81$, and an average axis ratio of $\langle b/a\rangle_{i'}\!=\!0.69$.

Comparing the central and outer region in terms of mean structural parameters and nucleation fraction, we see some interesting differences mostly in the mean magnitudes and effective radii.~Non-nucleated dwarfs in the central region are brighter and larger with $\langle M_{i'}\rangle\!=\!-11.99 \pm 0.12$\,mag and $\langle r_{{\rm eff},i'}\rangle\!=\!0.61 \pm 0.03$\,kpc, relative to non-nucleated dwarfs in the outer region which have $\langle M_{i'}\rangle\!=\!-11.65 \pm 0.10$\,mag and $\langle r_{{\rm eff},i'}\rangle\!=\!0.55\pm0.02$\,kpc.~Nucleated dwarfs in the central region have significantly fainter average luminosities but are similar in average size compared to their nucleated counterparts in the outer footprint, $\langle M_{i'}\rangle\!=\!-12.43\pm0.21$\,mag,  $\langle r_{{\rm eff},i'}\rangle\!=\!0.91 \pm 0.04$\,kpc (central) and $\langle M_{i'}\rangle\!=\!-13.87 \pm 0.2$\,mag and $\langle r_{{\rm eff},i'}\rangle\!=\!0.95 \pm 0.04$\,kpc (outer).~These differences are more pronounced when comparing the nucleated and non-nucleated dwarf population, i.e.~the non-nucleated dwarf population is fainter than nucleated dwarfs by $\Delta\langle M_{i'}\rangle\!=\!0.44$\,mag in the central region and $\Delta\langle M_{i'}\rangle\!=\!2.25$\,mag in the outer region.~In addition, the non-nucleated dwarf population is, on average, smaller than the nucleated dwarf population by $\Delta\langle r_{{\rm eff},i'}\rangle\!=\!0.3$\,kpc and $\Delta\langle r_{{\rm eff},i'}\rangle\!=\!0.4$\,kpc, in the central and outer region, respectively.

Non-nucleated dwarfs have similar mean sizes and luminosities independent of local environmental density, i.e.~central vs.~outer region.~However, central nucleated dwarfs are on average about $1.4$ magnitudes fainter than nucleated dwarfs in the outer regions.~In the central regions nucleated and non-nucleated have similar magnitudes but different mean sizes.~However, in the outer region, nucleated dwarfs are on average brighter and larger than the non-nucleated dwarfs.~Table~\ref{tab:dwarftable} lists the IDs, coordinates, and structural parameters for the dwarf candidates at cluster-centric radii $r_{\rm NGC1399}\!>\!r_{\rm vir}/4$, complementing the sample from \cite{eigen18} for the inner region ($r_{\rm NGC1399}\!<\!r_{\rm vir}/4$).

\begin{figure*}
\centering
%trim=left bottom right top
\includegraphics[trim=0.3cm 0.5cm 0.3cm 0.3cm,clip,width=0.9\textwidth]{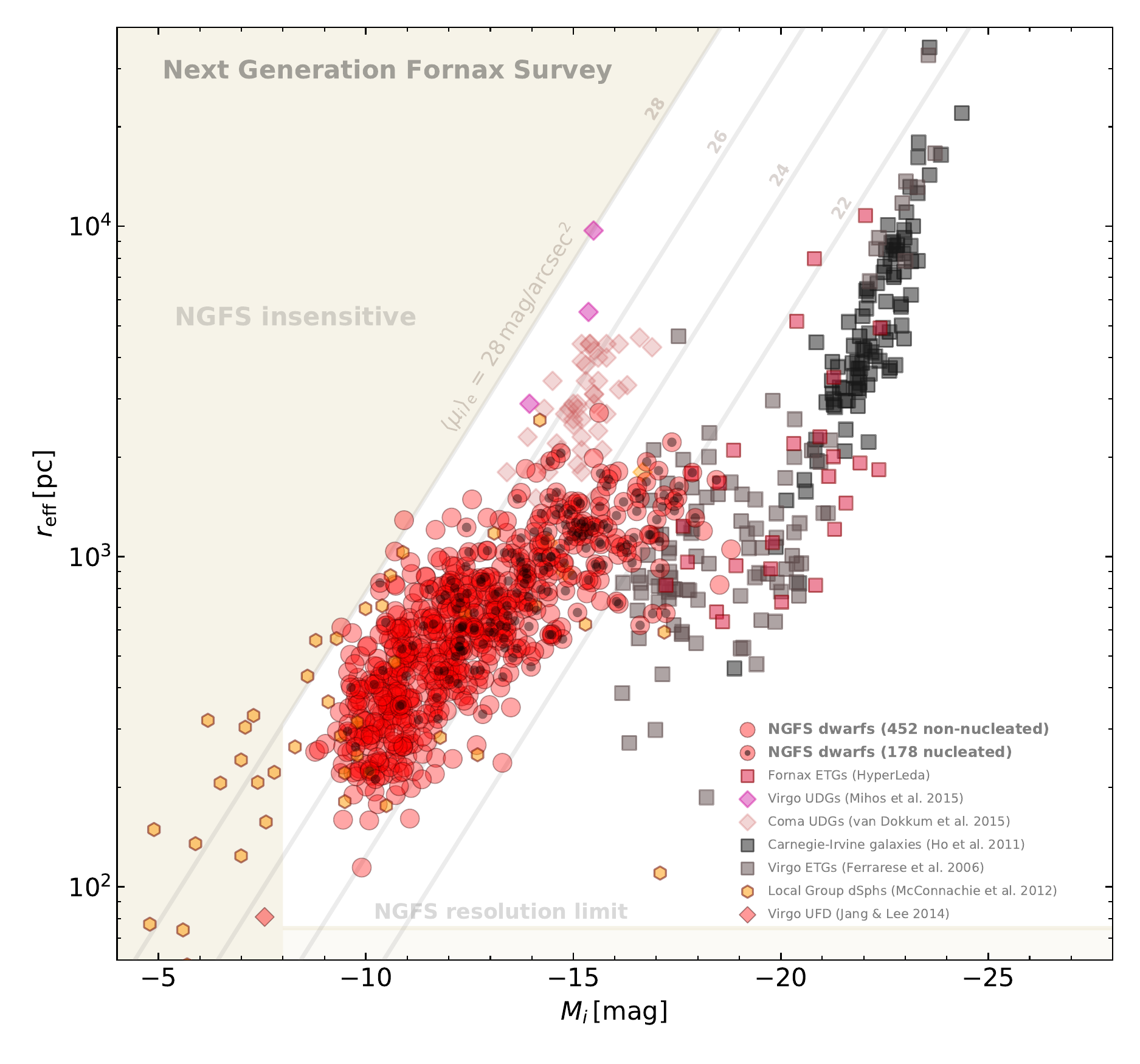}
\caption{Size-luminosity relation for dwarf and giant galaxies in Fornax and the nearby universe (see the legend at the bottom right).~Galaxy size is given by the effective radius and the luminosity represented as the absolute $i'$-band magnitude.~The NGFS dwarf population are shown in red circles with black dots indicating the nucleated galaxies.~The data for bright Fornax ETGs were taken from the HyperLeda database ({\footnotesize \url{http://leda.univ-lyon1.fr}}), where the effective radii were computed, in order of preference, from \cite{caon94}, \cite{deVauc91}, and \cite{lau89}.~Lines of constant average effective surface brightness are indicated for $\langle\mu_{i'}\rangle_e\!=\!28,26,24,22$\,mag arcsec$^{-2}$.~An approximate surface brightness limit of our NGFS data is illustrated by the shaded region toward the top left.~The spatial resolution limit ($0.8\arcsec$ in the $i'$ filter, corresponding to $\sim\!78$\,pc) is indicated by light shaded region in the bottom part of the diagram, below this limit objects appear as unresolved point sources in our NGFS data at the distance of Fornax \citep[$m\!-\!M\!=\!31.51$\,mag or $D_L\!=\!20.0$\,Mpc,][]{bla09}.}
\label{fig:dwarfsequence}
\end{figure*}

%%%%%%%%%%%%%%%%%%%%%%%%%%%%%%%%%%%%
%%%%%%%%%%%%%%%%%%%%%%%%%%%%%%%%%%%%
%%%%%%%%%%%%%%%%%%%%%%%%%%%%%%%%%%%%

\section{Discussion}
\label{sec:disc}

\subsection{Size-Luminosity Relation}
Scaling relations are useful tools to gain insight into the link of the formation processes between different astronomical objects.~We illustrate in Figure~\ref{fig:dwarfsequence} the size-luminosity relation in terms of the effective radius and absolute {\it i'}-band magnitude for the entire NGFS dwarf sample.~A total of 452 non-nucleated and 178 nucleated dwarfs are shown.~Thirteen dwarfs do not have structural parameter information due to very low surface brightness and/or complicated contamination in their nearby environment (e.g.~bright star spikes, detector blemishes, crowded field).~To map a large luminosity range, we overplot different galaxy samples, including Local Group dwarfs \citep{McCon12}, ultra-diffuse galaxies from the Coma and Virgo clusters \citep{vanDokkum15, Mihos15}, and giant ellipticals from Fornax, Virgo and the Carnegie-Irvine catalog \citep{Ferr06, Ho11}.~The NGFS dwarf sample by itself covers a range in absolute magnitude of $-18.80\!\le\!M_{i'}\!\le\!-8.78$\,mag and effective radii $r_{{\rm eff},i'}\!=\!0.11\!-\!2.72$\,kpc, comprising an effective surface brightness from $\langle\mu_{i'}\rangle_e\!=\!20\!-\!28$\,mag\,arcsec$^{-2}$.~The sequence of giant ellipticals stretching from the upper-to-center-right of the diagram is connected to that of the NGFS dwarfs and dSphs (center to lower-left) by an intermediate bridge of galaxies.~The bridge spans the $-20\!<\!M_{i'}\!<\!-15$\,mag, $0.6 \le r_{\rm eff,i'}/{\rm kpc}\le 2$ parameter space and blends the brightest NGFS dwarfs with the faint regime of ETGs.~We note here that the dwarfs in this group consist primarily of nucleated candidates such that 44/63 of these dwarfs show clear nuclei.

Ultra-diffuse galaxies (UDGs) seem to follow their own sequence, roughly along constant effective surface brightness, avoiding the bridge between dwarf and giant galaxies.~Although UDGs seem to have similar magnitudes to the brightest dwarf galaxies they are much more extended.~UDGs have been detected in multiple environments \citep[e.g.][]{Mihos15, koda15, mun15, vanDokkum15, jan17, lee17, ven17}.~They show signatures of massive dark-matter halos \citep[e.g.][]{bea16, vanDokkum16} and their population size scales with the mass of the central halo \citep{vdB16,vdB17, jan17}.~Together with predictions from theoretical studies this suggests that UDGs are a consequence of rapidly spinning, massive halos ($\ga\!10^{10}\,M_\odot$) that recently fell into denser environments \citep[e.g.][]{ron17, amo18}.~In our NGFS footprint, six UDG candidates are found in the central region, and one in the outer region.~Their magnitudes and effective radii are in the range $-15.62\!\le\!M_{i'}\!\le\!-13.85$\,mag and $r_{{\rm eff},i'}\!=\!1.79\!-\!2.72$\,kpc, respectively.~Of the seven candidates, two harbor a nuclear star cluster.~Their properties make them very similar to UDGs found in other galaxy cluster environments, such as Coma and Virgo \citep{vanDokkum15, Mihos15}.~A curious Local Group counterpart is the currently disrupting Sagittarius dwarf galaxy \citep[$M_{V}\!=\!-13.5$\,mag and $r_{\rm eff}\!=\!2.6$\,kpc,][]{McCon12} with its nucleus and the central star cluster M54 \citep[][]{bel08, muc17}.

Figure~\ref{fig:dwarfsequence} shows that the nucleation fraction of the NGFS dwarf sample decreases strongly with luminosity.~The overall nucleation fraction is 28\% for the entire luminosity range of the NGFS dwarfs.~Nonetheless, the nucleation fraction reaches $\sim\!90$\% at bright magnitudes, i.e.~$M_{i'}\la-17$\,mag, and drops to zero at the faintest galaxy luminosities, i.e.~$M_{i'}\ga-9.56$\,mag.~This limit marks the currently faintest nucleated galaxy in the NGFS dwarf galaxy sample \citep[see also][]{mun15,yob18}.

\subsection{Spatial Distribution}

\begin{figure*}[!t]
\centering
\includegraphics[width=0.9\textwidth]{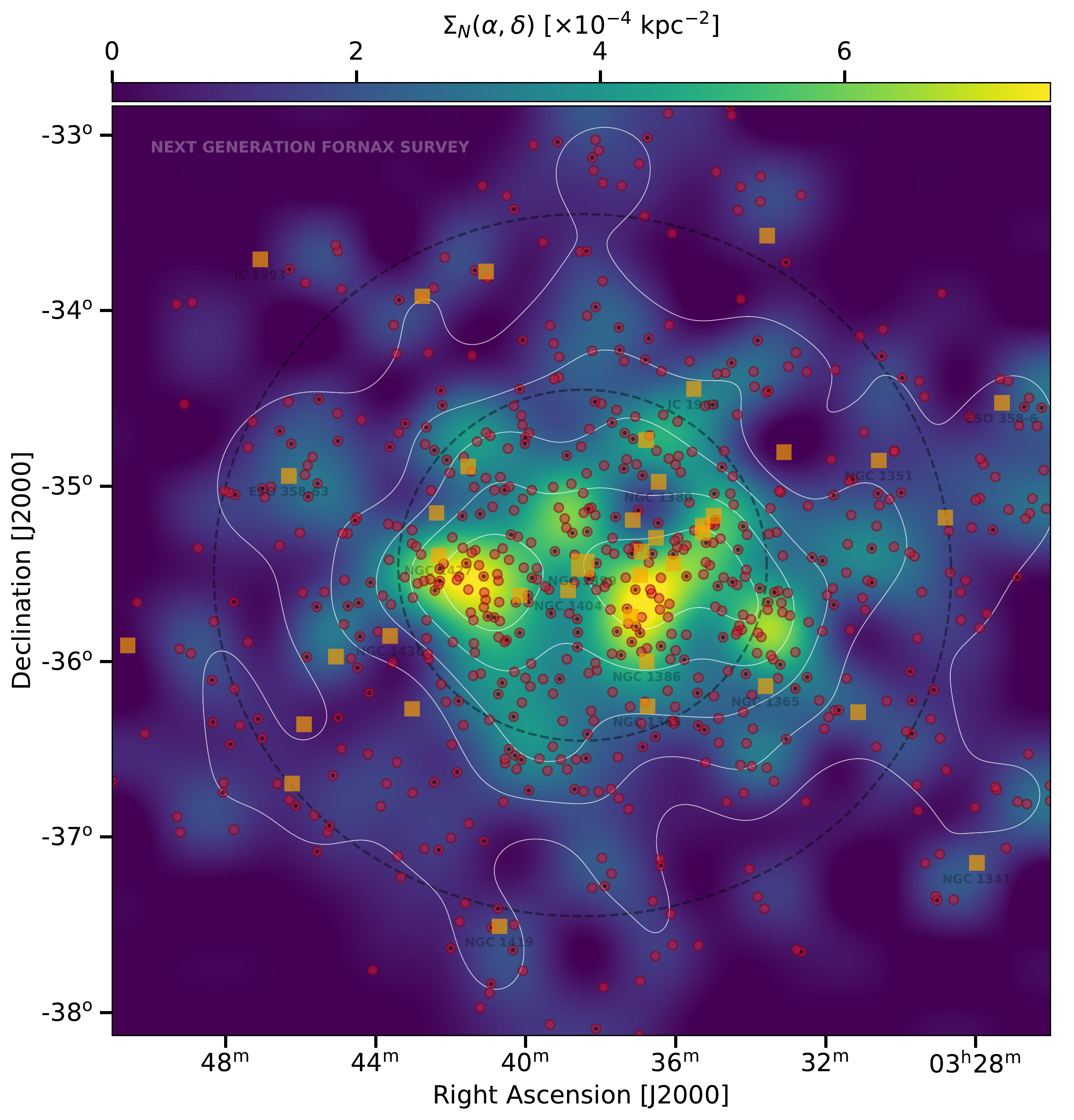}
\caption{The distribution of dwarf galaxies (fainter than $M_{i'}\!\approx\!-19$\,mag) in the Fornax galaxy cluster within about half of its virial radius ($\sim\!r_{\rm vir}/2$) shown as a surface number density distribution (see colorbar scale).~Overplotted are the NGFS dwarf candidates (red dots) and bright Fornax galaxies ($M_{i'}\!\lesssim\!-19$\,mag, orange squares) with NGC\,1399 shown by the large square.~Black dashed circles show NGC\,1399-centric distances of $r_{\rm vir}/4$ ($\simeq\!350$\,kpc) and $r_{\rm vir}/2$ ($\simeq\!700$\,kpc), while solid contours represent 2D Gaussian KDEs with a 0.25 degree kernel width.}
\label{fig:spatial_hists}
\end{figure*}

\begin{figure*}[!t]
\centering
\includegraphics[width=0.475\linewidth]{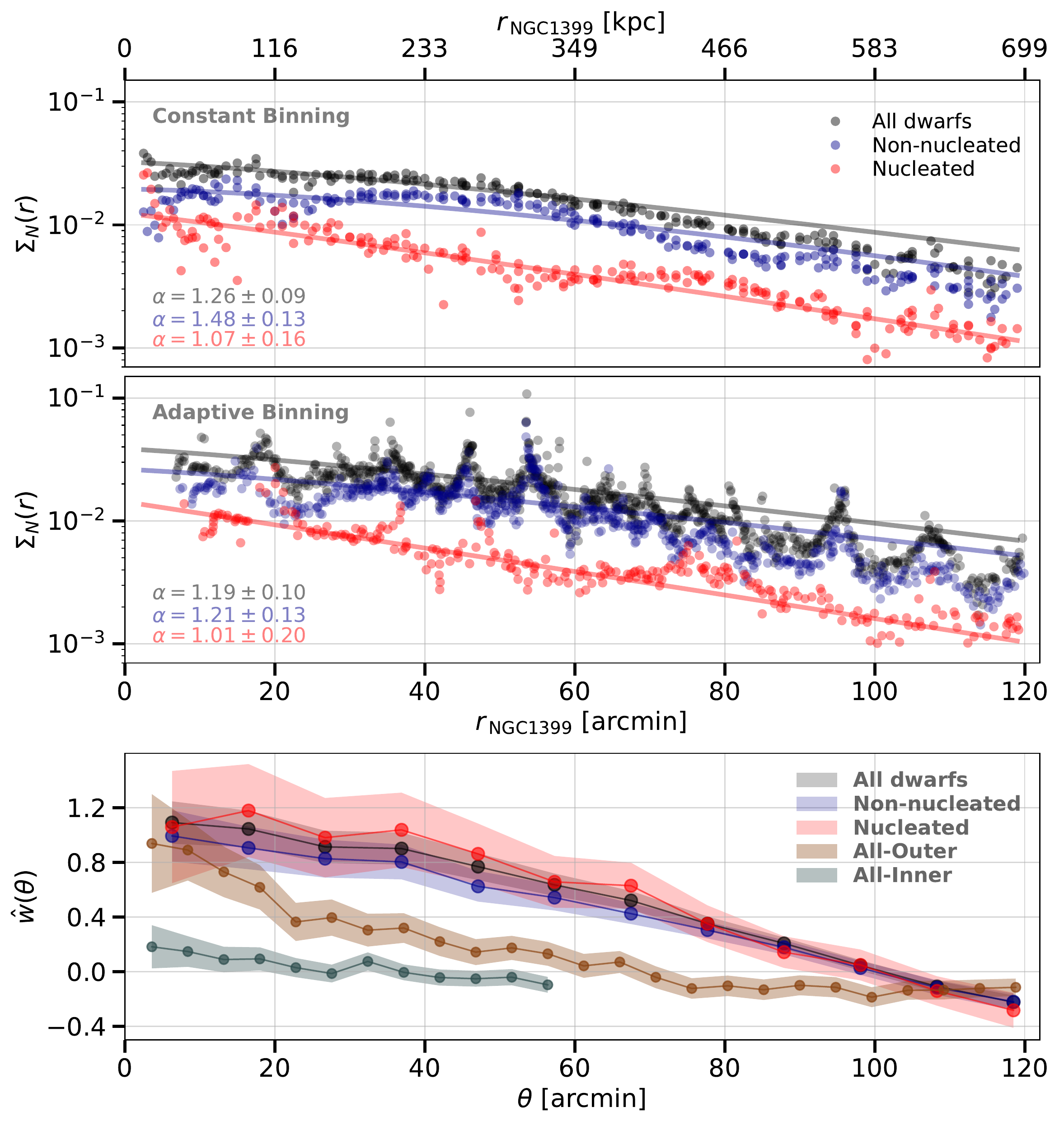}
\includegraphics[width=0.519\linewidth]{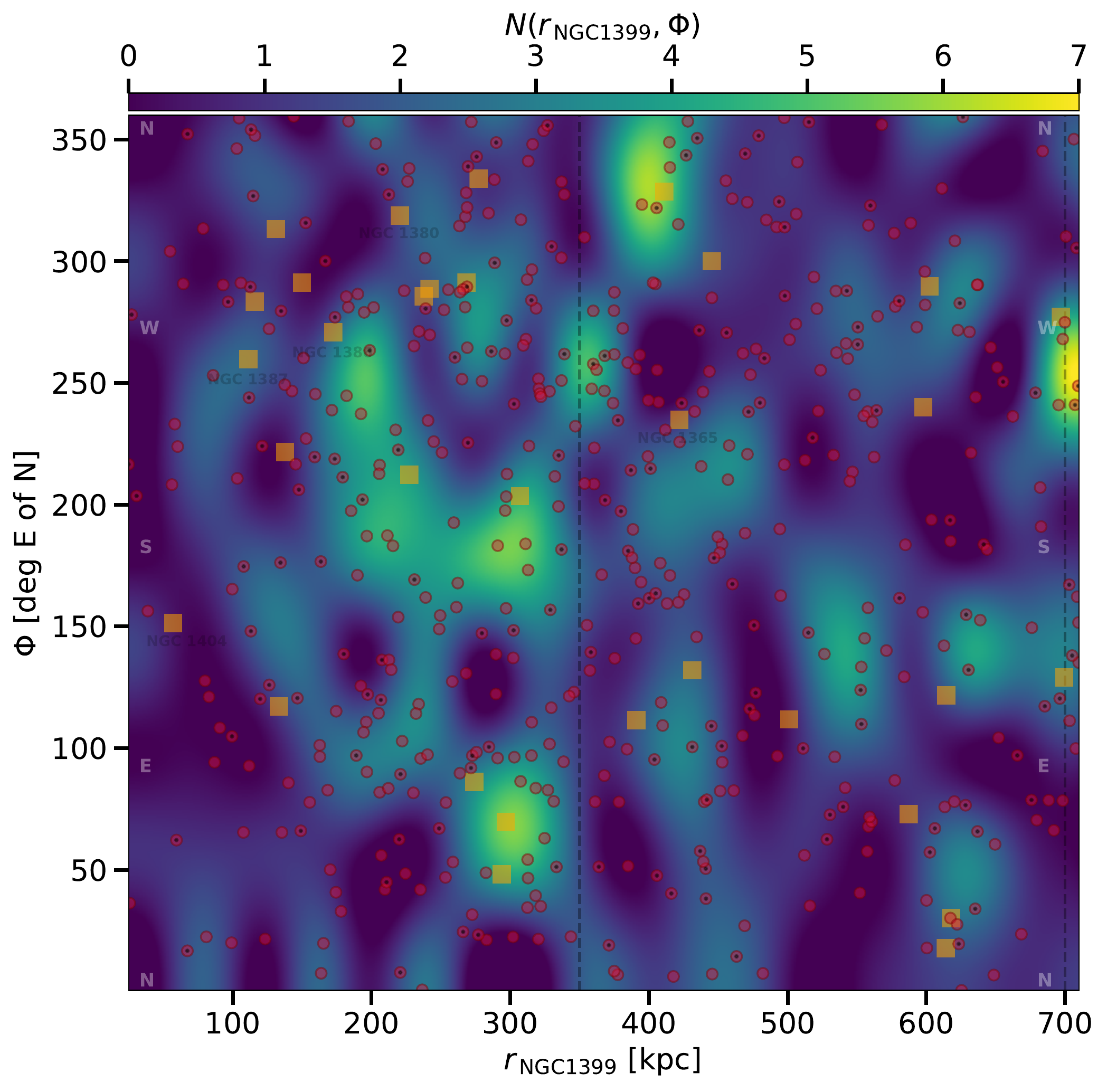}
\caption{The projected surface number density and clustering profiles for NGFS dwarf galaxy candidates.~{\it Left, top panels}:\,The two upper panels show radial surface number density distributions of dwarfs using constant (top) and adaptive (middle) binning strategies (see text).~An Einasto profile \citep[][]{ein89} is fit to each of these distributions, with the exponential slope and bootstrapped uncertainties indicated.~Only the homogeneously sampled region within $700$\,kpc ($r_{\rm vir}/2$) of NGC\,1399 is considered.~{\it Left, bottom panel}:\,The two-point angular correlation functions and associated bootstrapped uncertainties are shown for the total (blue), and outer/inner samples (brown/grey; split at $r\!=\!350$\,kpc).~{\it Right panel}:\,2D density distribution showing dwarf clustering in the $\Phi-r_{\rm NGC1399}$ space, smoothed using Lanczos interpolation.~Red circles are dwarf galaxies while orange squares show the locations of bright Fornax galaxies.~Dashed vertical lines indicate the NGC\,1399-centric radii of $r_{\rm vir}/4$ ($\simeq\!350$\,kpc) and $r_{\rm vir}/2$ ($\simeq\!700$\,kpc).
}
\label{fig:spatial_profs}
\end{figure*}

Figure~\ref{fig:spatial_hists} shows the spatial distribution of the dwarf galaxy candidates in the NGFS survey region, with dashed black circles indicating NGC\,1399-centric radii of $r_{\rm vir}/4$ and $r_{\rm vir}/2$ that correspond to $\sim\!350$ and $700$\,kpc at the distance of Fornax.~The dwarfs and giant galaxies are distributed throughout the field with the projected dwarf galaxy surface number density profile, \surfdens, shown by the color shading, computed along a $15\times15$ bins grid---corresponding to physical bin sizes of $152\!\times\!146$\,kpc$^2$---and show the resulting 2D histogram smoothed with Lanczos interpolation.~We also estimate \surfdens\ by a non-parametric kernel density estimate using a Gaussian kernel of 0.25\,deg bandwidth and show resulting curves of iso-density contours by grey-scaled thin solid lines.~We point out that variations in surface brightness limits due to bright galaxy haloes do not affect the results since the typical size of a galaxy halo is negligible compared to the area studied here and the structures found therein.

Both \surfdens\ estimates show a general concentration of dwarfs in the core regions of Fornax within $r\!\la\!350$\,kpc.~NGC\,1399 itself can be seen to occupy an apparent saddle-point between two main dwarf galaxy over-densities towards the East and West \citep[see Figs.~\ref{fig:footprint} and \ref{fig:spatial_hists}, and also][]{mun15}.~The projected distribution of dwarfs in the Western over-density generally follows that of the giant galaxies, which may suggest a physical association.~While the current data cannot confirm such a connection, we note the contrast with the $\sim\!200$\,kpc-scale over-density to the East.~This group shows a more regular morphology, and lies in between to the projected positions of only two bright galaxies, complicating the notion of physical origins with nearby giant hosts.~Weaker density contrasts are found for a third ``orphan'' group of dwarf candidates located near $(\alpha,\delta)\approx(03^h33^m, -35.75^{\rm o})$.

The top two left-hand panels of Figure~\ref{fig:spatial_profs} show the \surfdens\ profile calculated with respect to NGC\,1399.~To guard against potential biases introduced in choosing arbitrary binsizes, we resample the dwarf galaxies using both a constant and an adaptive binning strategy.~For the former, we choose constant bins between 5$\arcmin$ and 25$\arcmin$, in steps of 1$\arcmin$.~At each step we calculate \surfdens\ in each of the corresponding annuli and show the aggregated data as dots.~We have done this procedure for non-nucleated and nucleated dwarfs, and the entire dwarf sample.~Similarly, the middle panel of Figure~\ref{fig:spatial_profs} shows the data based on adaptive bin sizes with the three samples described above.~Here, bins are chosen such that each of them contains exactly $N$ dwarf candidates with $5\leq N\leq25$.~The constant binning produces a much smoother variation in \surfdens, while the adaptive binning shows a sensitivity to local dwarf over-densities resulting in artificially small annuli and corresponding spikes in the \surfdens\ profile.~We approximate the dwarf \surfdens\ profiles by a power-law model, following other radially dependent projected densities \citep[e.g.~\surfdens$\propto \exp(A\,R^\alpha)$;][]{ein65, ser63} by Markov-Chain Monte Carlo (MCMC) sampling.~We use MCMC with a normal prior to sample the posterior probability of ($A$, $\alpha$) $10^5$ times, which allows sufficient burn-in to skip over stochastic or unreliable results from early iterations so that each chain converges to consistent, well-defined peaks in the marginalized parameter estimate distributions.~The resulting $\Sigma_N(r)$ estimates are shown in the two top left-panels of Figure~\ref{fig:spatial_profs} as solid curves.~The text displays the means of the posterior probability density functions for $\alpha$ alongside the corresponding 95\% confidence limits.~The projected radial surface number density for the non-nucleated dwarfs has a flat distribution up to $\sim\!350$\,kpc and slowly declines beyond that radial distance.~On the other hand, nucleated dwarfs have a steeper $\Sigma_N(r)$ distribution than non-nucleated dwarfs, being more concentrated in the inner regions near the cD galaxy NGC\,1399 and decreasing rapidly outwards.~This is similar to the results of \cite{lis07} who found based on SDSS photometry ($M_B\lesssim-13$\,mag) that the bright nucleated dwarf galaxy population in the Virgo galaxy cluster is more centrally concentrated than the non-nucleated dwarf population.

The bottom left-panel of Figure~\ref{fig:spatial_profs} shows the results of a two-point angular correlation function \citep[$\hat{w}(\theta)$;][]{lan93} analysis of the dwarf galaxy candidates, using the same three samples (non-nucleated, nucleated, and all dwarfs).~The two-point angular correlation function quantifies the excess probability of finding galaxy pairs at a given angular separation over a random distribution without a restriction to Gaussianity \citep[e.g.][]{connolly02, sat09}, and is typically used to constrain cosmological parameters and structure formation models \citep[e.g.][]{bernardeau02, coo02, teg02, teg04, dol06}.~Noting the steepening of the \surfdens\ profile at $\sim\!350$\,kpc and the $\Sigma_N(r)$ distribution, we split the total sample into those dwarf galaxies inside and outside of this radius and estimate $\hat{w}(\theta)$ for each sub-population.~The solid lines along with $\pm1\sigma$ bounds show the results for the three samples, in addition to all-outer and all-inner dwarfs, with correlation lengths $\theta$ binned in steps of 10\arcmin\ for the total, non-nucleated and nucleated sample, and 5$\arcmin$ for the two sub-samples.~Estimating $\hat{w}(\theta)$ using different bin sizes reveals mild deviations but with the overall behaviors unchanged.

Given that the likelihood of finding two points separated by an angular distance $\theta$ compared to a purely uniform distribution is encapsulated by $\hat{w}(\theta)$, we find particularly strong evidence for dwarf clustering on sub-$100$\,kpc scales for the overall and outer dwarf population.~In particular, the outer dwarf galaxies appear more likely to be clustering on scales approaching $\sim\!50$\,kpc with a notable decrease on scales $\ga\!100$\,kpc.~Overall, the apparent smaller clustering scale appears superimposed on the $\sim\!350$\,kpc-scale over-density shown in Figure~\ref{fig:spatial_hists}.~This larger clustering scale is reflected by the almost flat \surfdens\ profile within $\sim\!350$\,kpc, outside of which a general steepening of the profile is seen, modulo local, projected dwarf clustering and sub-groups.~Conversely, the dwarf population inside of $r\!\approx\!350$\,kpc does not appear to show as strong evidence for clustering at any scale, but we note that the limited spatial region will tend to mute $\hat{w}(\theta)$.~In any case, we check against an underlying uniform distribution of dwarf galaxies by creating a large artificial set of 3D dwarf galaxy coordinates, uniformly distributed within a 1\,Mpc radius sphere centered on NGC\,1399. We limit this population to those lying within the coordinate limits of our observed sample, and extract the 2D projected distances from NGC\,1399.~Two-sample Kolmogorov-Smirnov tests comparing the simulated NGC\,1399-centric projected separations and position angles to the new dwarfs rule out a flat surface density distribution of dwarf galaxies within the NGFS field of view at a very high confidence level (i.e.~$p\!=\!0.004$ for separations and $p\!\ll\!0.001$ for galaxy position angles).

The right-hand panel of Figure~\ref{fig:spatial_profs} shows an alternate view of the spatial distribution.~Here we show the projected number density as a function of azimuthal angle ($\Phi$; degrees East of North) and radial distance from NGC\,1399.~Symbols are as in Figure~\ref{fig:spatial_hists} with dashed lines indicating $r_{\rm vir}/2$ and $r_{\rm vir}/4$, which corresponds to 350 and 700\,kpc from NGC\,1399.~We apply $(\Delta\Phi\!=\!36^\circ,\Delta r_{\rm NGC1399})$ binning and smooth the 2D histogram with Lanczos interpolation, which serves to highlight dwarf groupings aligning along ``lines-of-sight'' to NGC\,1399, which do not appear as obvious as in the projected $(\alpha,\delta)$ space (see Fig.\,\ref{fig:spatial_hists}).~Nevertheless, similar $\la100$\,kpc-scale overdensities are apparent at all radii, further supporting the non-uniformity of the Fornax dwarf galaxy population, in particular the E-W bimodality shown near $r_{\rm NGC\,1399}=r_{\rm vir}/4 \!\approx\!350$\,kpc.~Spectroscopic observations are required to assess the phase-space coherence of the found dwarf galaxy overdensities.

Taken together, the evidence of clustering at $\la\!100$\,kpc scales within the central cluster-centric radius of $\sim\!1$\,Mpc of Fornax broadly concurs with the growing observational evidence for the common occurrence, and importance, of dwarf galaxy pairs and groups in low-mass galaxy evolution and transformation in the Local Universe \citep[e.g.][]{mar12, ann16, ord16, sti17}.~Theoretical considerations predict that close associations between dwarf galaxies should be common in that as many as 50\% of $10^6\,M_\odot$-scale dwarf galaxies might be expected to have companions within $\sim\!50$\,kpc \citep[][]{wet15, whe15}.~Given the isolated natures of many of the recently discovered dwarf pairs/groups, combined with the locations of other purported dwarf groups ranging out to $\sim100$\,kpc from giant galaxy hosts, the current findings may indicate that at least some of these groups are truly interacting in the halos of Fornax giant galaxy members, or that they may have origins in the primordial universe and are falling into the Fornax cluster environment for the first time.

%% If you wish to include an acknowledgments section in your paper,
%% separate it off from the body of the text using the \acknowledgments
%% command.
\acknowledgments
Y.O.-B.\ acknowledges financial support via CONICYT-Chile (grant CONICYT-PCHA/Doctorado Nacional/2014-21140651) and the DAAD through the PUC-HD Graduate Exchange Fellowship.~P.E.~acknowledges support from the Chinese Academy of Sciences (CAS) through CAS-CONICYT Postdoctoral Fellowship CAS150023 administered by the CAS South America Center for Astronomy (CASSACA) in Santiago, Chile.~T.H.P.~acknowledges the support through the FONDECYT Regular Project No.~1161817 and the BASAL Center for Astrophysics and Associated Technologies (PFB-06).~T.H.P. and A.L. gratefully acknowledge ECOS-Sud/CONICYT project C15U02. This project used data obtained with the Dark Energy Camera (DECam), which was constructed by the Dark Energy Survey (DES) collaboration.

%This project is supported by... This research has made use of the NASA Astrophysics Data System Bibliographic Services, the NASA Extragalactic Database, and the SIMBAD database and VizieR catalog access tool, operated at CDS, Strasbourg, France \citep{wen00}.

%% To help institutions obtain information on the effectiveness of their 
%% telescopes the AAS Journals has created a group of keywords for telescope 
%% facilities.
%
%% Following the acknowledgments section, use the following syntax and the
%% \facility{} or \facilities{} macros to list the keywords of facilities used 
%% in the research for the paper.  Each keyword is check against the master 
%% list during copy editing.  Individual instruments can be provided in 
%% parentheses, after the keyword, but they are not verified.

\vspace{1mm}
\facilities{CTIO:Blanco/DECam}

%% Similar to \facility{}, there is the optional \software command to allow 
%% authors a place to specify which programs were used during the creation of 
%% the manusscript. Authors should list each code and include either a
%% citation or url to the code inside ()s when available.

\software{astropy \citep{ast13},
	  matplotlib \citep{hun07},
	  scikit-learn, \citep{ped12},
	  astroML, \citep{van12}
	  SCAMP \citep{ber02},
	  SWARP \citep{ber06},  
          Source Extractor \citep{ber96},
          Galfit \citep{peng02}
          }

\end{document}